\documentclass{ifacconf}

\usepackage[T1]{fontenc}
\usepackage[utf8]{inputenc}
\usepackage[numbers,square]{natbib} 

\usepackage{graphicx}      
\usepackage{amsmath}
\usepackage{amssymb}
\usepackage{mathrsfs}
\usepackage{subfig}
\usepackage{pstricks}
\usepackage{psfrag}

\DeclareMathOperator*{\argmin}{arg\,min}

\DeclareMathOperator*{\Prob}{Prob} 
\newcommand{\topp}[0]{\textsf{T}}

\renewcommand{\Re}{\mathbb{R}}
\renewcommand{\paragraph}[1]{\smallskip\noindent\textbf{#1.} }
\newtheorem{pbm}{Problem}

\newcommand{\bbm}{[\begin{matrix}}
\newcommand{\eem}{\end{matrix}]}

\newcommand{\BM}{\left[\begin{matrix}}
\newcommand{\EM}{\end{matrix}\right]}

\begin{document}
\begin{frontmatter}

\title{Optimal control of discrete-time switched linear systems via continuous parameterization} 

\author[First]{Jérémie Kreiss} 
\author[Second]{Laurent Bako}
\author[Second]{Eric Blanco}

\address[First]{Laboratoire Ampère, INSA de Lyon, Université de Lyon  \\ 20, Avenue Albert Einstein, 69100 Villeurbanne, France
 \\E-mail: jeremie.kreiss@insa-lyon.fr}
\address[Second]{Laboratoire Ampère -- Ecole Centrale de Lyon -- Université de Lyon \\ 36, Avenue Guy de Collongue, 69134 Ecully, France}

\begin{abstract}                
The paper presents a novel method for designing an optimal controller
for discrete-time switched linear systems. The problem is formulated as
one of computing the discrete mode sequence and the continuous input
sequence that jointly minimize a quadratic performance index. State-of-
art methods for solving such a control problem suffer in general
from a high computational requirement due to the fact that an exponential
number of switching sequences must be explored. The method of this
paper addresses the challenge of the switching law design by introducing
auxiliary continuous input variables and then solving a non-smooth block-sparsity inducing 
 optimization problem.
\end{abstract}


\end{frontmatter}
\setcitestyle{numbers}
\section{Introduction}

Switched systems constitute a class of dynamic systems consisting of a finite number of subsystems which are  activated one after another over time  by a switching signal \citep{IntroductionToHS}, \citep{HandbookOfHSControl}, \citep{SupervisoryHybrifSystems}. In many cases, the switching signal is an external input which, together with the continuous input vector, can be used to control the behavior of the switched system. Examples of real processes that can be represented as switched systems with external discrete and/or continuous inputs are: autonomous vehicles,  chemical processes, electrical circuits, etc. (see e.g., \citep{IntroductionToHS} and \citep{HandbookOfHSControl} for a background). The problem of controlling switched systems of this type therefore involves designing a continuous control law along
with a switching sequence to achieve some performance specifications. We will be interested more specifically in optimal control design for discrete-time switched linear systems (SLS).  

The existing approaches to the problem of optimal control for switched systems can be classified into two groups: 
the ones addressing the continuous-time case \citep{Xu04-TAC,Deaecto11-Automatica,Riedinger13-TAC,Riedinger15-Automatica,Senger15-TAC} and those pertaining to the discrete-time case \citep{Zhang09-Automatica}, \citep{Gorges11-TAC}. Other classifications can be done with respect to the length of the control horizon (finite or infinite), the simplifying assumptions posed a priori on the structure of the discrete or the continuous input or the conceptual nature of the methods. One observation that can be made is the following. Considering the quadratic optimal control of SLS in continuous-time where no structure is imposed in advance on the discrete and continuous control policies, 
 there is so far no exact solution for both  finite and infinite control horizons. In discrete-time  an exact solution has been derived in  \citep{Zhang09-Automatica} for the case of finite control horizon. However direct numerical implementation proves to be so expensive that it is not affordable in practice. Therefore a relevant question of major importance is how to develop some suboptimal strategies that would be much less expensive while still being close to optimality. Some relaxations have been discussed in \citep{Zhang09-Automatica,Gorges11-TAC} for this purpose. However the resulting control algorithms still suffer from an exponential demand in storage capacity.


In this paper, we focus on the quadratic optimal control of switched linear  systems. The discussion here is restricted to finite-horizon problems but the envisioned ultimate goal is to extend it to infinite horizon. 
We first observe that the origin of the huge complexity associated with computing the solution to this optimization problem is the presence of discrete variables.  Therefore a key idea in our approach is to parameterize the discrete inputs by auxiliary control variables which are continuous. With this reformulation, the SLS optimal control problem becomes completely continuous but nonconvex. We then discuss some convex  relaxation strategies. More precisely, we propose solving a sequence of convex problems in order to estimate the auxiliary control variables. The whole process yields an  implementation which is shown in simulation to  coincide statistically very often with the true optimal control. The advantage of the proposed computational scheme is that it has only a polynomial complexity. Moreover it requires the same storing capacity (for the sequence of positive semidefinite matrices generated by the Riccati recursion) as the solution to the simpler linear quadratic optimal problem.  

The structure of this paper is as follows. In Section \ref{section2}, we formulate the switched quadratic control problem. We discuss the expression of the optimal solution and the associated complexity issue. 
In Section \ref{section3} we present the new continuous parameterization of the discrete control variable and  develop a  four-steps algorithm for solving it approximately. A numerical illustration is provided in Section \ref{section4}. Finally, some concluding remarks are given in section \ref{section5}.
\section{Problem Formulation}
\label{section2}

\subsection{Switched linear systems}
We consider a discrete-time switched linear system (SLS) described by
\begin{equation}\label{eq:SLS}
x(k+1)=A_{\sigma(k)}x(k)+B_{\sigma(k)}u(k), \quad x(0)=x_0
\end{equation}
where $k\in \mathbb{Z}_+$ is  a time index, $x(k)\in \Re^n$ and $u(k)\in \Re^{n_u}$ are respectively the state and the continuous input at time $k$, $x_0$ is the initial state; $\sigma(k)\in \Omega\triangleq \left\{1,\ldots,q\right\}$ refers to the value taken by the switching signal (also called here the discrete input) at time $k$.  $\Omega$ is a finite set collecting the indices of the different subsystems of the SLS \eqref{eq:SLS}. For any $i\in \Omega$, the pair $(A_i,B_i)\in \Re^{n\times n}\times \Re^{n\times n_u}$ of matrices is associated with the subsystem $i$. Throughout the paper, we use notations of the type  $u(\cdot)$ to designate the entire sequence $\left\{u(k)\right\}$. 

It is important to note that the switching signal $\sigma(\cdot)\triangleq \left\{\sigma(k)\right\}$ is viewed here as an external input. For simplicity, we assume that $\sigma(k)$ can be selected freely in $\Omega$ without any constraint. The control problem of interest is that of designing jointly a continuous control sequence $u(\cdot)$ and a discrete control sequence $\sigma(\cdot)$ so as to minimize a certain performance index over a finite time horizon. For this purpose it will be assumed throughout the paper that the SLS \eqref{eq:SLS} is stabilizable.  




\subsection{Switched optimal quadratic problem}
We consider a quadratic performance index  associated with system \eqref{eq:SLS} in the form 
\begin{equation}
\begin{array}{l}
J_0(x_0, u(\cdot),\sigma (\cdot))=\dfrac{1}{2}x^{\textsf{T}}(N)\Psi x(N)\\
\hfill+\dfrac{1}{2}\sum\limits_{\substack{k=0}}^{N-1} \big[x^{\textsf{T}}(k)Q(k)x(k)+u^{\textsf{T}}(k)R(k)u(k)\big]
\end{array}
\label{J1}
\end{equation}
where $N$ denotes the control horizon,  $\left\{R(k)\right\}_{k=0}^{N-1}\subset \Re^{n_u\times n_u}$ is a given sequence of symmetric positive definite matrices; $\left\{Q(k)\right\}_{k=0}^{N}\subset \Re^{n\times n}$ with $Q(N)=\Psi$ represents a sequence of positive semidefinite matrices. 
The problem of interest in this paper is stated as follows. 


\begin{pbm}\label{pbm:SLS-LQ} 
Given the performance matrices $\left\{Q(k)\right\}_{k=0}^{N}$ and $\left\{R(k)\right\}_{k=0}^{N-1}$, find a continuous input sequence $u(\cdot)$ and a discrete input sequence $\sigma(\cdot)$ that minimize the performance index \eqref{J1}   subject to the switched system equation \eqref{eq:SLS}.  In more formal terms, this is equivalent to solving the optimization problem 
\begin{equation}\label{eq:Opt-control}
\begin{aligned}
& \min_{u(\cdot),\sigma (\cdot)} J_0\big(x_0, u(\cdot),\sigma (\cdot)\big)\\ 
& \quad  \text{s.t.  Eq. \eqref{eq:SLS}.}
\end{aligned}
\end{equation}
\end{pbm}
We start by observing that the solution to problem \eqref{eq:Opt-control} can be mathematically characterized in a quite straightforward way using, for example, the Bellman optimality principle \citep{Bertsekas12-Book}. However a direct implementation of the optimal control suffers from an exponential complexity in both storing and computational resources. The goal of this paper is to discuss some  alternative formulations and corresponding solutions of the quadratic optimal control problem for SLS so as to yield more efficient implementations.  

Let us first  characterize the  solution to problem \eqref{eq:Opt-control}. 
For this purpose,  let $J_k(x,u(\cdot),\sigma(\cdot))$ denote the performance index corresponding to the situation when the system starts in an arbitrary state $x\in \Re^n$ at time $k$ and evolves under the action of the inputs $u(\cdot)$ and $\sigma(\cdot)$. Introduce the function $J_k^{*}:\Re^n\rightarrow \Re$ defined by 
\begin{equation}\label{eq:Cost-to-go}
\begin{aligned} J_k^{*}(x) &=
	\underset{\substack{u(k),\cdots,u(N-1) \\ \sigma(k),\cdots,\sigma(N-1)}}{\text{min}} \quad J_k(x,u(\cdot),\sigma(\cdot)) \\
	& \: \text{s.t. } x(t+1)=A_{\sigma(t)}x(t)+B_{\sigma(t)} u(t), \:  x(k)=x \\
	&  \qquad  \qquad \qquad \:  t=k,\ldots, N-1. 
\end{aligned}
\end{equation}
$J_k^{*}(x)$ is  called the \textit{value function} or the \textit{cost-to-go}. It is interesting to note from \eqref{J1} that  $J_N^{*}(x)=x^{\textsf{T}}\Psi x$ for all $x\in \Re^n$.
Equipped with the notation \eqref{eq:Cost-to-go}, the Bellman principle of optimality \citep{Bertsekas12-Book} can be expressed as 
\begin{equation}
J_k^{*}(x)=\min_{\substack{(u,i) \in \mathbb{R}^{n_u}\times \Omega}}\big[\ell(x,u,k)+J^*_{k+1}(A_ix+B_iu)\big]
\label{eq:Bellman}
\end{equation}
where $\ell(x,u,k)=\dfrac{1}{2}\left[x^{\textsf{T}}Q(k)x+u^{\textsf{T}}R(k)u\right]$ is the running cost. 
\begin{thm}
Consider the control problem \eqref{eq:Opt-control} and let the functions $J_k^*$ be defined as in \eqref{eq:Cost-to-go}. Denote with $u^*(\cdot)$, $\sigma^*(\cdot)$, $x^*(\cdot)$ respectively the optimal continuous input, discrete input and continuous state. 
Then the following statements hold. 
\begin{enumerate}
\item[1.] The value function $J_k^{*}$ is quadratic and can be written as 
\begin{equation}\label{induction hypothesis}
J_{k}^{*}(x)=x^{\textsf{T}}P^*(k) x \quad \forall \: x\in \Re^n,\: \:  \forall k \in \left\{0,\ldots, N\right\}, 
\end{equation}
where $\left\{P^*(k)\right\}$ is a sequence of matrices generated recursively backward in time according to the Ricatti recursion defined by 
\begin{equation}\label{eq:Ricatti-sequence}
	P^*(k)=\rho_{\sigma^*(k),k}(P^*(k+1)), \quad  P^*(N)=\Psi
\end{equation}
with 
\begin{equation}\label{eq:Ricatti-Map}
\begin{aligned}
		\rho_{i,k}(P)=&Q(k)+A_i^{\textsf{T}}PA_i\\
		&-A_i^{\textsf{T}}PB_i\big(R(k)+B_i^{\textsf{T}}PB_i\big)^{-1}B_i^{\textsf{T}}PA_i.  
\end{aligned}
\end{equation}

\item[2.] The optimal continuous input is given by
\begin{equation}\label{uopt}
u^{*}(k) = -K_{\sigma^{*}(k),k}\big(P^*(k+1)\big)x^*(k) 
\end{equation}
where
$
 K_{i,k}(P)=\big( R(k)+ B_{i}^{\textsf{T}}PB_{i}\big)^{-1} B_i^{\textsf{T}}PA_{i}. 
$
\item[3.] The optimal discrete input $\sigma^*(\cdot)$ is given by
\begin{equation}\label{sigmaopt}
\sigma^{*}(k)\in \underset{i\in \Omega}{\text{argmin}}\big[x^*(k)^{\textsf{T}}\rho_{i,k}(P^*(k+1))x^*(k)\big].
\end{equation}
\end{enumerate}

\end{thm} 

\begin{pf}
The proof follows by a simple backward induction exploiting the Bellman optimality equation \eqref{eq:Bellman}. It is therefore omitted. 
\end{pf}

\subsection{Implementation of the optimal control}
In order to implement the control law \eqref{uopt}-\eqref{sigmaopt}, we need to fully compute offline the  sequence $\{P^*(k)\}_{k=0}^N$ by the Ricatti recursion \eqref{eq:Ricatti-sequence}. Note that this recursion depends  on the optimal switching signal $\sigma^*(\cdot)$ which in turn depends on the optimal continuous state $x^*(\cdot)$ as can be seen from Eq. \eqref{sigmaopt}. Unfortunately the optimal state is not available offline. This is a source of a major difficulty.  
To cope with this challenge an elegant solution has been developed in  \citep{Zhang09-Automatica}.   The authors first showed that the value function given in \eqref{induction hypothesis} can be re-expressed as 
$$J_k^*(x)=\min_{P\in \mathcal{H}_{k}} x^\topp Px $$ 
where $\left\{\mathcal{H}_k\right\}$ is a sequence of sets of symmetric positive semidefinite matrices defined by\footnote{Note that in the setting of \citep{Zhang09-Automatica} the weighting matrices $Q$ and $R$ depend on the subsystem index, not on time. As a consequence $\rho$ is indexed there only by $i$.}
\begin{equation}\label{Hk}
\left\{\begin{aligned}
&\mathcal{H}_N=\{\Psi\} \quad \\
&\mathcal{H}_k =\big\{\rho_{i,k}(P): P \in \mathcal{H}_{k+1}, i \in \Omega \big\}, \:  k =N-1,\ldots,0.
\end{aligned}\right.
\end{equation}
The sets $\left\{\mathcal{H}_k\right\}$ collect indeed all the possible values of the Ricatti sequence \eqref{eq:Ricatti-sequence} for all admissible switching signals. Since the definition of  the sets $\left\{\mathcal{H}_k\right\}$ is now freed from the dependence on the continuous state, they can be computed offline and stored. Once this is done the optimal control law can be obtained online
by applying 
\begin{equation}\label{eq:online-sigma-opt}
	\sigma^*(k)\in\argmin_{i\in\Omega, P\in \mathcal{H}_{k+1}}x^*(k)^\top \rho_{i,k}(P)x^*(k) 
\end{equation}
along with \eqref{uopt}. 

It turns out that the trick of \citep{Zhang09-Automatica} makes the optimal control \eqref{uopt}-\eqref{sigmaopt} implementable. This is done however at the price of a  huge complexity. In effect, the cardinality of the sets $\mathcal{H}_k$ grows exponentially fast with respect to the control horizon $N$. For example, the cardinality of $\mathcal{H}_k$ is about $q^{N-k}$. Indeed the above implementation requires  an exponential load in terms of both computational and storage resources. This complexity affects all the steps of the implementation: offline computation and storage of the sets $\left\{\mathcal{H}_k\right\}$ and online reading and search for the optimal discrete control by Eq. \eqref{eq:online-sigma-opt} over the sets $\left\{\mathcal{H}_k\right\}$. 
To reduce the complexity, a suboptimal solution is discussed in \citep{Zhang09-Automatica}. But it is fair to observe that some shortcomings still persist. First, the proposed procedure does not alleviate the off-line computational load ; it only allows for a saving of the necessary storage capacity. Moreover the  cardinality reduction algorithm has still to test all the elements in the sets $\mathcal{H}_k$ hence resulting in an exponential complexity.  

This complexity restricts the applicability of the solution of \citep{Zhang09-Automatica} to the control of switched systems with small number of subsystems and small control horizons. Noting that exponential complexity is generated by the presence of discrete variables in the optimization problem, we discuss here a new approach which relies on a continuous parameterization of the switching sequence. 

\section{Proposed solution}\label{section3}
The proposed design method relies on two main ideas:
\begin{itemize}
	\item A continuous parameterization of the discrete control variable $\sigma(\cdot)$. This consists in replacing the discrete input in the SLS \eqref{eq:SLS} with continuous variables called auxiliary control variables. Consequently, problem \eqref{eq:Opt-control} can be written as a constrained optimization problem in only continuous variables hence getting rid of its combinatorial feature.
	\item A nonsmooth block-sparsity-inducing optimization involving the auxiliary control variables. The  purpose of this is to enforce their expected structure as will be described next. 
\end{itemize}
 
\subsection{Continuous parameterization of the discrete input}
A starting point is to notice that the SLS dynamics in \eqref{eq:SLS} can be written in the form
\begin{equation}\label{eq:parameterization}
x(k+1)=A_i x(k)+B_iu(k)+f_i(k), \: \forall i\in \Omega
\end{equation}
where $f_i(k)=(A_{\sigma(k)}-A_i)x(k)+(B_{\sigma(k)}-B_i)u(k)$. Each $f_i(k)$ can be interpreted as the difference between the state  of the SLS at time $k+1$ under $\sigma(k)$ and the state that would have been obtained if $\sigma(k)$ was set equal to $i$. It follows that for any time instant $k$, $f_{\sigma(k)}(k)=0$.  

From now on let us forget about the explicit expressions of the $f_i(k)$'s and view them just as unknown control variables satisfying the following constraint: for all $k$, there exists  $j=j(k)\in \Omega$ such that $f_{j}(k)=0_{n,1}$ where $0_{n,1}$ denotes a $n$-dimensional vector with all entries equal to zero. By letting $\bar{f}(k)=[\begin{matrix}f_1(k)^\topp & \cdots &f_q(k)^\topp \end{matrix}]^\topp \in \Re^{qn}$, the  above  constraint can be written as $\bar{f}(k)\in \mathcal{S}$ with $\mathcal{S}$ being a subset of $\Re^{qn}$ defined by 
\begin{equation}\label{eq:non-convex-constraint}
\begin{array}{l}
\mathcal{S}=\big\{v=(v_1^\textsf{T},\cdots,v_q^\textsf{T})^\textsf{T}\in \mathbb{R}^{qn}, \mbox{ with }v_i \in \mathbb{R}^n, i \in \Omega: \\
\hfill \exists j \in \Omega : v_j=\textbf{0}_{n,1} \big\}.
\end{array}
\end{equation}
There are many other equivalent ways of representing the set $\mathcal{S}$. One of those which are smooth is the following 
$$
\begin{aligned}
	\mathcal{S}=\Big\{v=(v_1^\textsf{T},\cdots,v_q^\textsf{T})^\textsf{T}\in \mathbb{R}^{qn}, \mbox{ with }&v_i \in \mathbb{R}^n, i \in \Omega: \Big.\\
	&\Big.\prod_{i=1}^q \left\|v_i\right\|=0\Big\}
\end{aligned}
$$
for any vector norm $\left\|\cdot\right\|$ on $\Re^n$. 
The  variable $\bar{f}(k)$ is then regarded as an auxiliary control variable to be computed along with the continuous input variable. 
Based on this parameterization we now restate the control problem as follows.
\begin{pbm}\label{pbm:Second-Formulation}
Given the matrices $\left\{Q(k)\right\}_{k=0}^{N}$ and $\left\{R(k)\right\}_{k=0}^{N-1}$, find a continuous input sequence $u(\cdot)$ and an  auxiliary input sequence $\bar{f}(\cdot)$ that minimize the performance index 
\begin{equation}\label{J2}
\begin{aligned}
	V(x_0,& u(\cdot),\bar{f} (\cdot) )=\dfrac{1}{2}x^{\textsf{T}}(N)\Psi x(N)+\sum_{k=0}^{N-1}\ell(x(k),u(k),k)\\
\end{aligned}
\end{equation}
subject to system \eqref{eq:parameterization} written in expanded form as 
\begin{equation}
\left[\begin{array}{c} x(k+1)\\ \vdots \\ x(k+1) \end{array}\right]=\left[\begin{array}{c} A_1\\ \vdots \\ A_q \end{array}\right]x(k)+\left[\begin{array}{c} B_1\\ \vdots \\ B_q \end{array}\right]u(k)+\bar{f}(k)
\label{new system}
\end{equation}
and the constraint $\bar{f}(k)\in \mathcal{S}$. 
\end{pbm}
To write this in a more compact form, define 
$$
\begin{aligned}
	&\bar{A}=\BM A_1^\topp & \cdots & A_q^\topp \EM^\topp, \quad\bar{B}=\BM B_1^\topp & \cdots & B_q^\topp \EM^\topp, \\
	&L_{n,q}=\BM I_n & \cdots I_n\EM ^\topp = \textbf{1}_q\otimes I_n,
\end{aligned}
$$
where $\textbf{1}_q=\bbm 1 & \cdots & 1\eem^\topp$ is a vector of $q$ ones, $I_n$ is the $n$-dimensional identity matrix and $\otimes$ denotes the Kronecker product.
Then problem \eqref{pbm:Second-Formulation} can be written as a constrained optimization problem  as follows
\begin{equation}\label{eq:SLS-LQ-2}
\begin{aligned}
 \min_{u(\cdot),\bar{f} (\cdot)} &V\big(x_0, u(\cdot),\bar{f} (\cdot) \big)\\ 
& \text{s.t. } \: L_{n,q} x(k+1)=\bar{A} x(k) + \bar{B} u(k)+\bar{f}(k)\\
&  \qquad  \bar{f}(k)\in \mathcal{S}, \: k=0, \ldots,N-1\\
& \qquad  x(0)=x_0.
\end{aligned}
\end{equation}
It is important to note that problems \eqref{eq:Opt-control} and \eqref{eq:SLS-LQ-2} are equivalent. But contrary to \eqref{eq:Opt-control}, problem \eqref{eq:SLS-LQ-2} is an optimization problem in which all the decision variables are continuous. One can therefore hope for designing an algorithm that solves it in polynomial time. A remaining challenge  to deal with is the non convexity of \eqref{eq:SLS-LQ-2}.  


\subsection{An algorithm in four stages} 
We now ask the question of how to compute numerically the solution to problem \eqref{eq:SLS-LQ-2}. 
Because of the non-convex constraint $\bar{f}(k)\in \mathcal{S}$, the problem is not convex. For efficiency of solving we therefore need to find a   convex relaxation. We will discuss a block-sparsity inducing optimization technique for that. 
The global computational procedure can be decomposed into four stages which are described next.


\subsubsection{Stage (a)}
In this first stage, the sequence $\left\{\bar{f}(k)\right\}_{k=0}^{N-1}$ is considered as a (unknown) parameter.  Hence the criterion $V$ is minimized with respect to $u(\cdot)$ only, i.e., we solve 
\begin{equation}\label{eq:Problem-StageA}
\begin{aligned}
 \kern-1em \min_{u(\cdot)} & \: V\big(x_0, u(\cdot),\bar{f} (\cdot) \big)\\ 
 \text{s.t. } & \: L_{n,q} x(k+1)=\bar{A} x(k) + \bar{B} u(k)+\bar{f}(k), \\
&  k=0, \ldots,N-1\\
&  x(0)=x_0,
\end{aligned}
\end{equation}
where the constraint on $\left\{\bar{f}(k)\right\}$ has been removed. This is a   convex program. One difference with the classical linear quadratic control problem is that the dynamic matrix $\bar{A}$  in \eqref{eq:Problem-StageA} is rectangular rather than square. Also the state $x(k+1)$ is repeated here $q$ times on left hand side of the dynamics equation.  Denote with $V_1(x_0,\bar{f}(\cdot))$ the resulting optimal functional, i.e.,  $V_1(x_0,\bar{f}(\cdot))=\min_{u(\cdot)} V\big(x_0, u(\cdot),\bar{f} (\cdot) \big)$.

Using the method of Lagrange multipliers, construct the Lagrangian $\bar{V}$ by embedding the constraints of \eqref{eq:Problem-StageA} in the cost functional. This yields
\begin{equation}
\begin{array}{l}
\bar{V}=V\big(x_0, u(\cdot),\bar{f}(\cdot)\big)+\\
\sum_{k=0}^{N-1}\bar{\lambda}^{\textsf{T}}(k+1) \big(\bar{A}x(k)+\bar{B}u(k)+\bar{f}(k)- L_{n,q} x(k+1)\big)
\end{array}
\label{extended cost}
\end{equation}
where $\left\{\bar{\lambda}(k)\right\}$, with  $\bar{\lambda}(k)=\bbm \lambda_1(k)^\topp & \cdots & \lambda_q(k)^\topp \eem^\topp$, is a sequence of Lagrange multipliers. Introduce now the discrete time Hamiltonian $H$ associated with the system \eqref{new system} and the performance index \eqref{J2} defined by
\begin{equation}
\begin{aligned}
H_k&=H\big(x(k),u(k),\bar{\lambda}(k+1),k)\\
&=\ell(x(k),u(k),k)+\lambda^{\textsf{T}}(k+1)\left(\bar{A}x(k)+\bar{B}u(k)+\bar{f}(k)\right)
\end{aligned}
\end{equation}
With this short-hand notation the extended cost \eqref{extended cost} takes the form 
\begin{equation}
\begin{aligned}
\bar{V}=\dfrac{1}{2}x(N)^{\textsf{T}}&\Psi x(N)-\bar{\lambda}^{\textsf{T}}(N)L_{n,q} x(N)\\
&+\sum\limits_{\substack{k=1}}^{N-1} \big(H_k-\bar{\lambda}^{\textsf{T}}(k)L_{n,q} x(k) \big) +H_0. 
\end{aligned}
\label{extended cost 2}
\end{equation}
Considering the minimization of $\bar{V}$ with respect to $u$, let us look at the effect of an elementary variation  $du$ of $u$ and of the induced change $dx$ in the state. These together induce a variation $d\bar{V}$ of $\bar{V}$, expressible by 
\begin{equation}
\begin{aligned}
& d\bar{V}= \left[\Psi x(N)-L_{n,q}^{\textsf{T}}\bar{\lambda}(N)\right]^{\textsf{T}}dx(N)\\
&+\sum_{\substack{k=1}}^{N-1}\left(\left[\dfrac{\partial H_k}{\partial x(k)}- L_{n,q}^{\textsf{T}}\bar{\lambda}(k)\right]^{\textsf{T}}dx(k)+\dfrac{\partial H_k}{\partial u(k)}du(k)\right)\\
& +\dfrac{\partial H_0}{\partial x(0)}dx(0)+\dfrac{\partial H_0}{\partial u(0)}du(0). 
\end{aligned}
\end{equation}
Note that  $dx(0)=0$ since the initial state is fixed. At the optimal value of $\bar{V}$ we must have $d\bar{V}=0$ as a consequence of any variation $du$ of the input and any subsequent variation $dx$ of the state. 
This implies that all terms in the previous equation must be set to zero.  
Thus for all $k =1,\ldots,N-1$,  $\bar{\lambda}(k)$ must satisfy 
\begin{equation}\label{eq:Adjoint-State}
L_{n,q}^{\textsf{T}}\bar{\lambda}(k)=\dfrac{\partial H_k}{\partial x(k)}, \quad  L_{n,q}^{\textsf{T}}\bar{\lambda}(N)=\Psi x(N). 
\end{equation}
Similarly, we must impose 
$$
\dfrac{\partial H_k}{\partial u(k)}=0.
\label{H sur u}
$$
It follows that the optimal continuous input can be expressed as
\begin{equation}
u(k)=-R^{-1}\bar{B}^{\textsf{T}}\bar{\lambda}(k+1).
\label{commande}
\end{equation}
We can rewrite \eqref{new system} and \eqref{eq:Adjoint-State}  respectively as follows
\begin{equation}
L_{n,q} x(k+1) = \bar{A}x(k)-\bar{B}R^{-1}\bar{B}^{\textsf{T}}\bar{\lambda} (k+1)+\bar{f}(k)\\
\label{state equation}
\end{equation}
with initial condition $x(0)=x_0$ and 
\begin{equation}\label{adjoint equation}
L_{n,q}^{\textsf{T}}\bar{\lambda}(k) = Qx(k)+\bar{A}^{\textsf{T}}\bar{\lambda}(k+1), 
\end{equation}
with the final constraint $L_{n,q}^{\textsf{T}}\bar{\lambda}(N) = \Psi x(N)$. 

Eqs \eqref{state equation} and \eqref{adjoint equation} form a system of linear  equations that characterizes completely the solution to \eqref{eq:Problem-StageA}.

\subsubsection{Stage (b)}
Now, we seek the minimal value of $V_1(x_0,\bar{f}(\cdot))$ with respect to $\bar{f}(\cdot)$. Since for any $k$ the solution $\bar{f}(k)$ is expected  to live in the set $\mathcal{S}$ defined in \eqref{eq:non-convex-constraint}, the optimization problem is as follows
\begin{equation}
\begin{aligned}
& \min_{\bar{f}(\cdot)}\: V_1\big(x_0,\bar{f}(\cdot)\big)\\
& \quad \text{s.t. } \:  \bar{f}(k) \in \mathcal{S} ~\forall k \in \{0,\cdots, N-1\}. 
\end{aligned}
\label{stage-b-non-convex-problem}
\end{equation}
As mentioned earlier, this problem is not convex. In order to keep polynomial complexity, we need to find a convex relaxation of it.  

To do so, let us  denote with  $\chi_{\mathcal{S}}:\Re^{nq}\rightarrow \Re\cup\left\{+\infty\right\} $ the indicator function of $\mathcal{S}$ defined from $\Re^{nq}$ to real extended line by $\chi_{\mathcal{S}}(z)=0$ if $z\in\mathcal{S}$ and $\chi_{\mathcal{S}}(z)=+\infty$ otherwise. Then \eqref{stage-b-non-convex-problem} is equivalent
$$
\min_{\bar{f}(\cdot)}\:\Big[ V_1\big(x_0,\bar{f}(\cdot)\big)+\sum_{k=0}^{N-1} \chi_{\mathcal{S}}(\bar{f}(k))\Big].
$$
To find a convex relaxation of the terms  $\chi_{\mathcal{S}}(\bar{f}(k))$, we view each vector $\bar{f}(k)$ as being relatively block-sparse in the sense that it must admit at least one subvector $f_j(k)$ which is equal to zero. With this in mind we replace $\chi_{\mathcal{S}}(\bar{f}(k))$ with a nonsmooth convex function, $\sum_{i=1}^{q}w_i(k)\left\|f_i(k)\right\|_2$ where the $w_i(k)$'s denote some positive weights. As is suggested by a certain number of results (see e.g., \citep{Bako16-Automatica,Bako13-SCL}), minimizing such a function enjoys the nice property that it is able to promote block sparsity hence yielding  a vector $\bar{f}(k)$ with some of its subvectors $f_i(k)$ potentially equal or close to zero.  The role of the weights $\left\{w_i(k)\right\}$ is to discriminate between the different subvectors of $\bar{f}(k)$. They can be selected, for example, iteratively as follows: solve the problem with all weights set to one and based on the resulting solution, retune the weights  through a simple rule of the form $w_i(k)=1/(\|\hat{f}_i(k)\|_2+\epsilon)$ for some small number $\epsilon>0$. For better numerical stability, one can consider normalizing a posteriori the weights along the $i$ dimension.  

Finally we formulate the following convex optimization problem: 
\begin{equation}
\!\!\!\! \begin{aligned}
&\min_{\bar{f}(\cdot)} \: \Big[\gamma_1 V_1\big(x_0,\bar{f}(\cdot)\big)+\gamma_2\sum_{k=0}^{N-1}\sum_{i=1}^{q}w_i(k)\left\|f_i(k)\right\|_2\Big].
\end{aligned}
\label{problem-stage-b}
\end{equation}

For writing simplicity, we have not given here the explicit expression of $V_1(x_0,f(\cdot))$. The derivation of such an expression follows from straightforward algebraic calculations departing from the system of linear equations  \eqref{state equation}-\eqref{adjoint equation}. 

\subsubsection{Stage (c)}
Let $\big\{\hat{\bar{f}}(k)\big\}$ with $\hat{\bar{f}}(k)=\bbm \hat{f}_1^\top(k)& \cdots & \hat{f}_q^\top(k)\eem^\top$ denote a solution to problem \eqref{problem-stage-b}. Since this is obtained under some relaxation of the initial problem, there is no guarantee that $\hat{\bar{f}}(k)$ will lie in the set $\mathcal{S}$ for all $k$. 
Hence, we need to project those $\hat{\bar{f}}(k)$ onto $\mathcal{S}$ in order to determine the switching sequence. The projection we used consists in selecting the discrete input to be the index $i\in \Omega$ such that  $\hat{f}_i$ has the  minimum norm among all, i.e.,
\begin{equation}
\hat{\sigma}(k)\in \underset{i \in \Omega}{\text{argmin}}\|\hat{f}_i(k)\|, \quad k =0,\ldots, N-1, 
\label{projection}
\end{equation}
with $\left\|\cdot\right\|$ denoting the vector $2$-norm. 
In some sense, this corresponds to forcing   to zero the $\hat{f}_i$ having the minimum norm.


\subsubsection{Stage (d)}
Once we have computed the switching sequence $\{\hat{\sigma}(k)\}_{k=0}^{N-1}$ offline, the SLS optimal control problem reduces to that of a linear time-varying system with matrices defined by $(A_{\hat{\sigma}(k)},B_{\hat{\sigma}(k)})$. The  solution of such a problem with respect to the quadratic cost \eqref{J1} can be determined as in the conventional case \citep{OptimalControl}. To obtain it, we can just apply \eqref{eq:Ricatti-sequence}-\eqref{eq:Ricatti-Map} with $\sigma^*(\cdot)$ replaced by $\hat{\sigma}(\cdot)$ to generate offline a sequence of matrices $\{\hat{P}(k)\}$. 
By storing this single Ricatti sequence, the final continuous and discrete inputs are then, similarly  as in \eqref{uopt}-\eqref{sigmaopt}, selected online as
\begin{align}
&\hat{u}(k)=-K_{\hat{\sigma}(k),k}(\hat{P}(k+1))\hat{x}(k) \label{eq:uhat}\\
&\hat{\sigma}(k)\in \argmin_{i\in \Omega}\big[\hat{x}(k)^{\textsf{T}}\rho_{i,k}(\hat{P}(k+1))\hat{x}(k)\big] \label{eq:sigma-hat}
\end{align}
with $\hat{x}(\cdot)$ denoting the associated state trajectory. This means that the  discrete input is recalculated online. 

As it turns out, the implementation proposed in this paper has a polynomial complexity. Moreover it requires storing only a single sequence of matrices $\{\hat{P}(k)\}$ just as in the solution of the linear quadratic problem (single linear subsystem). In comparison with \citep{Zhang09-Automatica}, the gain on the memory demand is significant. However in its current version, the proposed implementation is not guaranteed to yield the optimal control.  In view of the applied control policy \eqref{eq:uhat}-\eqref{eq:sigma-hat}, the sole objective of the procedure described above for solving Problem \ref{pbm:Second-Formulation} is the computation of the Ricatti sequence $\{\hat{P}(k)\}$.  


\section{Results}\label{section4}

Note that for all the following tests, the weights $\gamma_1$ and $\gamma_2$ are taken equals to $1$.
\subsection{A statistical test}\label{subsec:statistics}
In this section we challenge the capability of the proposed method to obtain the solution to the optimal quadratic control problem for switched systems of the form \eqref{eq:SLS}. For this purpose $100$ examples of two-dimensional switched systems, with each composed of $q=2$ subsystems, are generated at random using the MATLAB function \texttt{drss}. The initial state is also sampled from a Gaussian distribution $\mathcal{N}(0,20I_2)$. 
For each of these examples, the optimal control problem is that of finding the discrete and continuous inputs  to minimize a finite horizon performance index of the form \eqref{J1} with $N=15$ and with $Q(k)=I_2$ and $R(k)=1$ for all $k$.

For the sake of comparison, we also implement the optimal control law as described in \citep{Zhang09-Automatica} and recalled above in Eqs \eqref{Hk}-\eqref{eq:online-sigma-opt}. This is possible here since the number of subsystems and the control horizon are small. As a matter of fact the cardinality of $\mathcal{H}_0$ in \eqref{Hk} for the current experiment is about $2^{15}=32768$, which is computationally affordable on a standard computer. Let $J_i^{\text{opt}}$ and $J_i^o$, $i=1,\ldots,100$, denote respectively the optimal index and the value yielded by our method. 
For each of the $100$ examples, define the following relative error as an empirical measure of the distance to optimality
 $$\varepsilon_i=\dfrac{J_i^o-J_i^{\text{opt}}}{J_i^{\text{opt}}}.$$
The table below displays the distribution of $\varepsilon_i$ in terms of probabilities of the type $\Prob(\varepsilon_i\leq \alpha)$ with $\alpha$ being a threshold taking values in $\left\{10^{-10}, 10^{-8}, 10^{-7},10^{-5}\right\}$. It turns out that the solution achieved by the proposed method either coincides with the optimal index or  lies generally  in a small neighborhood of it. This suggests that for generic systems, the implemented strategy discussed in this paper has the potential of recovering the optimal control and this, at a much affordable price.  
\begin{table}[h!]
\renewcommand{\arraystretch}{1.4}
\begin{center}
\begin{tabular}{|c|c|c|c|c|c|}
\hline
Threshold on $\varepsilon_i$  & $10^{-5}$ & $10^{-7}$  & $10^{-8}$ & $10^{-10}$ & $0$\\ \hline
  $\%$ examples  & $100\%$ & $98\%$ &  $97\%$ & $96\%$ & $83\%$ \\ \hline
\end{tabular}
\caption{ }Distribution of the relative error $\varepsilon_i$ for the $100$ examples of switched systems: $(n,q)=(2,2)$.
\label{tab:distribution}
\end{center}
\end{table}

Repeating a similar experiment as above with this time examples of switched systems with state dimension $n=3$ and number of modes  $q=3$ yield the results reported in Table \ref{tab:distribution-3modes}. Note that for this last experiment the control horizon has been reduced to $N=10$ in order to alleviate the computational load associated with the computation of the exact optimal solution. A little degradation of the results can be observed in Table \ref{tab:distribution-3modes} in comparison with the results given in Table \ref{tab:distribution}. This may be due to numerical artefacts as a result of increased number of decision variables. The approximate performance index is still very close to the optimal one.   
\begin{table}[h!]
\renewcommand{\arraystretch}{1.4}
\begin{center}
\begin{tabular}{|c|c|c|c|c|c|c|}
\hline
Threshold  & $10^{-2}$ & $10^{-5}$ & $10^{-7}$  & $10^{-8}$ & $10^{-10}$ & $0$ \\ \hline
  $\%$ examples & $100\%$ & $96\%$ &  $93\%$ & $92\%$ & $90\%$ & $81\%$\\ \hline
\end{tabular}
\caption{ }Distribution of the relative error $\varepsilon_i$ for the $100$ examples of switched systems: $(n,q)=(3,3)$.
\label{tab:distribution-3modes}
\end{center}
\end{table}

\subsection{Illustration of performance on a single example}
\begin{figure}[hh]%
\psfrag{time}[][]{\scriptsize time}
\psfrag{sigma}[][]{\footnotesize $\sigma$}
\psfrag{state}[][]{\footnotesize State}
\psfrag{x1opt}[][]{\footnotesize $x_1^*$}
\psfrag{x2opt}[][]{\footnotesize $x_2^*$}
\psfrag{x1}[][]{\hspace{2pt}\footnotesize $\hat{x}_1$}
\psfrag{x2}[][]{\hspace{2pt}\footnotesize $\hat{x}_2$}
\subfloat[Discrete input: optimal $\sigma^*(\cdot)$ (red stars) and approximate one $\hat{\sigma}(\cdot)$ (blue circles). ]{\includegraphics[width=.97\columnwidth,height=3cm]{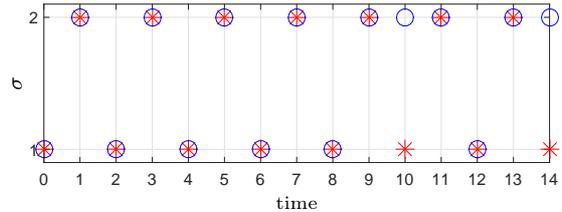}}\\
\subfloat[Continuous states $x^*(\cdot)$ and $\hat{x}(\cdot)$.]{\includegraphics[width=.97\columnwidth,height=4cm]{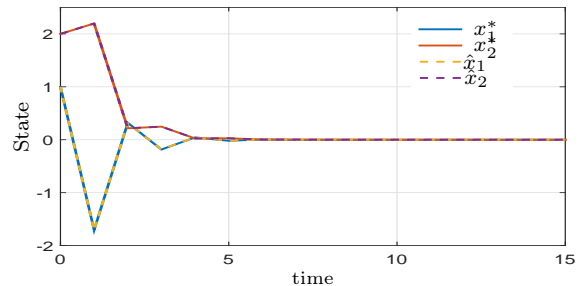}}%
\caption{Results obtained on the switched system defined by \eqref{eq:example}.}%
\label{fig:single-example}%
\end{figure}

 For illustration purpose, let us now focus on a single switched system.  The considered  example is in the form \eqref{eq:SLS} and consists of two linear subsystems with matrices defined by 
\begin{equation}\label{eq:example}
	\renewcommand{\arraystretch}{1.2}
	\begin{array}{ll}
	A_1=  \begin{bmatrix}0.9  & 0 \\  0.5  & 1.5\end{bmatrix},& B_1=\begin{bmatrix}2\\1 \end{bmatrix},\\
	A_2 = \begin{bmatrix} 1.1 &  1 \\ 0  & 0.8\end{bmatrix},& B_2=\begin{bmatrix}0\\1\end{bmatrix}.
	\end{array}
\end{equation}
It can be observed that none of the individual subsystems is stable. 
In this experiment, the control horizon is set to $N=15$; the initial state is $x_0=[\begin{matrix}1 & 2\end{matrix}]^\top$ and the weighting matrices of the performance index are defined as in Section \ref{subsec:statistics}. 
Applying the proposed method on this example yields the results presented in Figure \ref{fig:single-example}. It turns out that the obtained discrete input is equal to the optimal one except at the two time instants $t=10$ and $t=14$. However the impact of this difference is negligible on the performance index $J_0$ since we still get a relative error of $4.03\times 10^{-9}$. This is because the errors occur at a time when the state has almost already converged to zero as shown by Figure \ref{fig:single-example}-(b). We can even conjecture that small amplitude of the state makes it difficult to recover the optimal discrete input by \eqref{eq:sigma-hat}.   

\section{Conclusion}
\label{section5}
In this paper, we studied the discrete-time quadratic optimal control problem for switched systems on a finite time horizon. Based on a continuous parameterization of the discrete input, we proposed an approach that is able to yield the optimal solution in polynomial time with respect to the length of control horizon. Moreover  the proposed algorithm appears to be cheaper than most existing methods in terms of computational load and storage resources. Future research will focus on analyzing the properties of this method.  


\begin{thebibliography}{14}
\providecommand{\natexlab}[1]{#1}
\providecommand{\url}[1]{\texttt{#1}}
\providecommand{\urlprefix}{URL }
\expandafter\ifx\csname urlstyle\endcsname\relax
  \providecommand{\doi}[1]{doi:\discretionary{}{}{}#1}\else
  \providecommand{\doi}{doi:\discretionary{}{}{}\begingroup
  \urlstyle{rm}\Url}\fi

\bibitem[{Anderson and Moore(1990)}]{OptimalControl}
Anderson, B. and Moore, J. (1990).
\newblock \emph{Optimal Control : Linear Quadratic Methods}.
\newblock Prentice-Hall International.

\bibitem[{Bako and Lecoeuche(2013)}]{Bako13-SCL}
Bako, L. and Lecoeuche, S. (2013).
\newblock A sparse optimization approach to state observer design for switched
  linear systems.
\newblock \emph{Systems \& Control Letters}, 62, 143--151.

\bibitem[{Bako and Ohlsson(2016)}]{Bako16-Automatica}
Bako, L. and Ohlsson, H. (2016).
\newblock Analysis of a nonsmooth optimization approach to robust estimation.
\newblock \emph{Automatica}, 66, 132--145.

\bibitem[{Bertsekas(2012)}]{Bertsekas12-Book}
Bertsekas, D.P. (2012).
\newblock \emph{Dynamic Programming and Optimal Control}.
\newblock Athena Scientific.

\bibitem[{Deaecto et~al.(2011)Deaecto, Geromel, and
  Daafouz}]{Deaecto11-Automatica}
Deaecto, G.S., Geromel, J.C., and Daafouz, J. (2011).
\newblock Dynamic output feedback hinf control of switched linear systems.
\newblock \emph{Automatica}, 47, 1713--1720.

\bibitem[{G\"{o}rges et~al.(2011)G\"{o}rges, Iz\'{a}k, and Liu}]{Gorges11-TAC}
G\"{o}rges, D., Iz\'{a}k, M., and Liu, S. (2011).
\newblock Optimal control and scheduling of switched systems.
\newblock \emph{IEEE Transactions on Automatic Control}.

\bibitem[{Lemmon et~al.(1999)Lemmon, He, and
  Markovsky}]{SupervisoryHybrifSystems}
Lemmon, M.D., He, K.X., and Markovsky, I. (1999).
\newblock Supervisory hybrid systems.
\newblock \emph{IEEE Control Systems}, 19, 42--55.

\bibitem[{Lunze and Lamnabhi-Lagarrigue(2009)}]{HandbookOfHSControl}
Lunze, J. and Lamnabhi-Lagarrigue, F. (eds.) (2009).
\newblock \emph{Handbook of Hybrid Systems Control, Theory, Tools,
  Application}.
\newblock Cambridge University Press.

\bibitem[{Riedinger(2013)}]{Riedinger13-TAC}
Riedinger, P. (2013).
\newblock A switched {LQ} regulator design in continuous time.
\newblock \emph{IEEE Transactions on Automatic Control}, 59, 1322--1328.

\bibitem[{Riedinger and Vivalda(2015)}]{Riedinger15-Automatica}
Riedinger, P. and Vivalda, J.C. (2015).
\newblock Dynamic output feedback for switched linear systems based on a {LQG}
  design.
\newblock \emph{Automatica}, 54, 235--245.

\bibitem[{Senger and Trofino(2015)}]{Senger15-TAC}
Senger, G.A. and Trofino, A. (2015).
\newblock Switching rule design for affine switched systems with guaranteed
  cost and uncertain equilibrium condition.
\newblock \emph{IEEE Transactions on Automatic Control}, (To appear).

\bibitem[{Van Der~Shaft and Schumacher(1999)}]{IntroductionToHS}
Van Der~Shaft, A. and Schumacher, H. (1999).
\newblock \emph{An Introduction to Hybrid Dynamical Systems}.
\newblock Springer-Verlag London.

\bibitem[{Xu and Antsaklis(2004)}]{Xu04-TAC}
Xu, X. and Antsaklis, P.J. (2004).
\newblock Optimal control of switched systems based on parameterization of the
  switching instants.
\newblock \emph{IEEE Transactions on automatic control}, 49, 2--16.

\bibitem[{Zhang et~al.(2009)Zhang, Abate, Hu, and Vitus}]{Zhang09-Automatica}
Zhang, W., Abate, A., Hu, J., and Vitus, M. (2009).
\newblock Exponential stabilization of discrete-time switched linear systems.
\newblock \emph{Automatica}, 45, 2526--2536.

\end{thebibliography}

\end{document}